\documentclass[12pt]{article}
\usepackage{esfconf}
 \usepackage{epsfig}

\global\arraycolsep=2pt
\begin{document}


\title{Dimensional and Dynamical
Aspects of the Casimir Effect: Understanding the Reality and
Significance of Vacuum Energy}


\authors{Kimball A. Milton}


\addresses{Department of Physics and Astronomy, The University
of Oklahoma, Norman, OK 73019-0225 USA}


\maketitle


\begin{abstract}
Zero-point fluctuations in quantum fields give rise to
observable forces between material bodies, the so-called Casimir forces.  
In this lecture I present some results of
the theory of the Casimir effect, primarily
formulated in terms of Green's functions. There is an intimate relation
between the Casimir effect and van der Waals forces.
Applications to conductors
and dielectric bodies of various shapes will be given for the cases of
scalar, electromagnetic, and fermionic fields.  The dimensional dependence 
of the effect will be described. Finally, we ask the question: Is there
a connection between the Casimir effect and the phenomenon of
sonoluminescence?
\end{abstract}



\section{Introduction}

We may identify the zero-point energy of a system of quantum fields
 with the vacuum expectation value of the field energy,
\begin{equation}
{1\over2}\sum_a\hbar\omega_a=\int(d{\bf x})\langle T^{00}({\bf x})\rangle.
\end{equation}
In the vacuum both sides of this equality are
divergent and meaningless.  What is observable is the
{\em change\/} in the zero-point energy when matter is introduced.  In this
way we can calculate the Casimir forces.  
For a massless scalar field, the canonical energy-momentum tensor 
is\footnote{The ambiguity in defining the stress tensor is without effect.
For example, the `new-improved' traceless stress tensor gives the same
Casimir energy.}
\begin{equation}
T^{\mu\nu}=\partial^\mu\phi\partial^\nu\phi-{1\over2}g^{\mu\nu}\partial^\lambda
\phi\partial_\lambda\phi.
\label{stresstensor}
\end{equation}
The vacuum expectation value may be obtained 
 by taking derivatives of the casual Green's
function:
\begin{equation}
G({\bf x},t;{\bf x'},t')={i\over\hbar}\langle{\rm T}
\phi({\bf x},t)\phi({\bf x'},t')\rangle.
\label{vevgreen}
\end{equation}

Alternatively, we can calculate the stress on the material bodies.
Consider the original geometry considered by Casimir,
where he calculated the quantum fluctuation force between parallel,
perfectly conducting plates separated by a distance $a$ \cite{1}.
The force per unit area $f$ on one of the plates is given in terms of the
normal-normal component of the stress tensor,
\begin{equation}
f=\langle T_{zz}\rangle,
\end{equation}
For electromagnetic fields, the relevant stress tensor component is
\begin{equation}
T_{zz}={1\over2}(H_\perp^2-H_z^2+E_\perp^2-E_z^2).
\end{equation}
We impose classical boundary conditions on the surfaces,
\begin{equation}
H_z=0,\quad {\bf E}_\perp=0,
\end{equation}
and the calculation of the vacuum expectation value of the field components
reduces to finding the classical TE and TM Green's functions.  In general,
one further has to subtract off the stress that the formalism would give
if the plates were not present, the so-called volume stress,
and then the result of a simple calculation, which is sketched below,
is
\begin{equation}
f=\left[T_{zz}-T_{zz}({\rm vol})\right]=-{\pi^2\over 240 a^4}
\label{casimirsforce}
\hbar c,
\end{equation}
an attractive force.

The dependence on the plate separation 
$a$ is, of course, completely determined by dimensional
considerations.  Numerically, the result is quite small,
\begin{equation}
f=-8.11\,{\rm MeV}\,{\rm fm}\,a^{-4}
=-1.30\times 10^{-27} {\rm N}\,{\rm m}^2
\, a^{-4},
\end{equation}
and will be overwhelmed by electrostatic repulsion between the plates
if each plate has an excess electron surface density $n$ greater than $1/a^2$,
from which it is clear that the experiment must be performed at the
$\mu$m level.
Nevertheless, over twenty years, many attempts to measure this force
directly were made \cite{earlycasexp}.
(The cited measurements include insulators as well as conducting surfaces.)
Until recently, the most convincing experimental evidence came
from the study of thin helium
 films \cite{sa}; there the
corresponding Lifshitz theory \cite{lifshitz}
has been confirmed over nearly
 5 orders of magnitude in the van der Waals potential
(nearly two orders of magnitude in distance).  Quite recently, the Casimir 
effect between conductors has been confirmed at the 5\% level by
Lamoreaux \cite{lamoreaux},
and to perhaps 1\% by Mohideen and Roy \cite{mr}.  (In order to achieve the
stated accuracy, corrections for finite conductivity, surface distortions,
and perhaps temperature must be included.  For a brief review see 
Ref.~\cite{corr}.)

\section{Dimensional Dependence}
\subsection{Parallel Plates}

Here we wish to concentrate on dimensional dependence.
For simplicity we consider a massless scalar field $\phi$
confined between two parallel
plates in $d+1$ spatial dimensions separated by a distance $a$.  
Assume the field satisfies Dirichlet boundary conditions on the plates,
that is 
\begin{equation}
\phi(0)=\phi(a)=0.
\label{eq:bc}
\end{equation}
The Casimir force between the plates results from the zero-point energy
per unit ($d$-dimen\-sional) transverse area
\begin{equation}
u={1\over2}\sum \hbar\omega={1\over2}\sum_{n=1}^\infty\int{d^dk\over(2\pi)^d}
\sqrt{k^2+{n^2\pi^2\over a^2}},
\label{zeropt2}
\end{equation}
where we have set $\hbar=c=1$, and introduced normal modes labeled by
the positive integer $n$ and the transverse momentum $k$.
This may be easily evaluated by introducing a proper-time representation
for the square root, and by analytically continuing from negative $d$
we obtain
\begin{equation}
u=-{1\over2^{d+2}\pi^{d/2+1}}{1\over a^{d+1}}\Gamma\left(1+{d\over2}\right)
\zeta(2+d).\label{scalarenergy}
\end{equation}
which reduces to the familiar Casimir result at $d=2$:
\begin{equation}
u=-{\pi^2\over1440}{1\over a^3},\quad 
f_s=-{\partial\over\partial a}u=-{\pi^2\over 480}{1\over a^4}.
\label{casimir}
\end{equation}
This is, as expected, 1/2 of the electromagnetic result (\ref{casimirsforce}).

\begin{figure}
\centerline{
\epsfig{figure=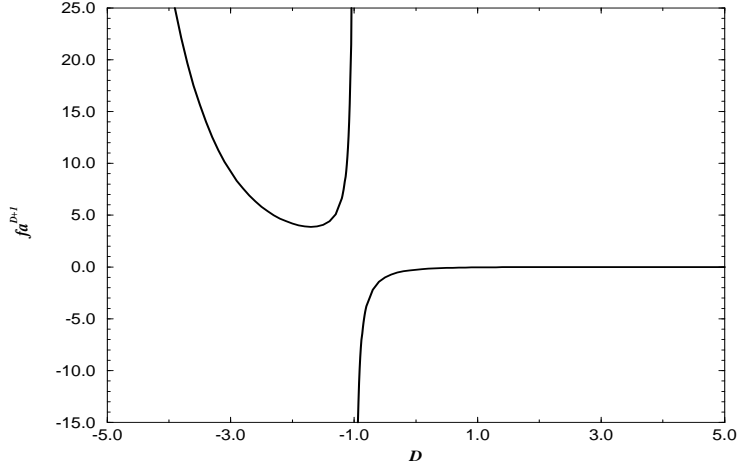,height=4.5in,width=3in,angle=270}}
\caption{A plot of the Casimir force per unit area $f$ in
Eq.~(\protect\ref{scalarforce})
for $-5<D<5$ for the case of a slab geometry (two parallel plates).
Here $D=d+1$.}
\label{fig2}
\end{figure}

This less-than-rigorous calculation can be put on a firm footing by a Green's
function technique.  Define a reduced Green's function by
\begin{equation}
G(x,x')=\int{d^d k\over(2\pi)^d}e^{i{\bf k}\cdot({\bf x-x'})}
\int{d\omega\over2\pi}e^{-i\omega(t-t')} g(z,z'),
\end{equation}
the (interior) solution of which vanishing at $z=0$, $a$, being
($\lambda^2=\omega^2-k^2$)
\begin{equation}
g(z,z')=-{1\over\lambda\sin\lambda a}\sin\lambda z_<\sin\lambda (z_>-a),
\label{scgreen}
\end{equation}
where $z_>$ ($z_<$) is the greater (lesser) of $z$ and $z'$.
The force per unit area on the surface $z=a$ 
is obtained by taking the discontinuity of the normal-normal component
of the stress tensor:
\begin{eqnarray}
f=\langle T_{zz}\rangle\bigg|_{z=z'=a-}-\langle T_{zz}\rangle\bigg|_{z=z'=a+}
=\int{d^dk\over(2\pi)^d}\int{d\omega\over2\pi}{\lambda\over2}(i\cot\lambda a
-1).
\end{eqnarray}
This is easily evaluated by doing a complex rotation in frequency:
$\omega\to i\zeta$:
\begin{equation}
f=-(d+1)2^{-d-2}\pi^{-d/2-1}{\Gamma\left(1+{d\over2}\right)\zeta(d+2)\over
a^{d+2}}.\label{scalarforce}
\end{equation}
Evidently, Eq.~(\ref{scalarforce}) is the negative derivative of the Casimir
energy (\ref{scalarenergy}) with respect to the separation between the plates:
$f=-{\partial u\over\partial a};$
this result has now been obtained by a completely well-defined approach,
so arguments about the conceptual validity of the Casimir effect are seen to
be without merit.
The force per unit area, Eq.~(\ref{scalarforce}), 
is plotted in Fig.~\ref{fig2}, where $a\to2a$ and $d=D-1$.

This general result was first derived by Ambj\o rn and Wolfram \cite{aw}.

\subsection{Fermion fluctuations}
The effect of massless fermionic fluctuations
between parallel plates embedded in three dimensional space, 
subject to ``bag model''  boundary conditions,
\begin{equation}
(1+i{\bf n}\cdot \mbox{\boldmath{$\gamma$}})G\bigg|_{z=0,a}=0,
\label{bmbc}
\end{equation}
 where $\bf n$ is the normal to the surface,
was first calculated by Ken Johnson \cite{johnson}.
In place of Eq.~(\ref{zeropt2}), the fermionic Casimir energy for $d=2$ is
formally
\begin{equation}
u_F=-2{1\over2}\sum_{n=0}^\infty
\int{d^2k\over(2\pi)^2}\sqrt{k^2+{(n+1/2)^2\pi^2\over a^2}},
\end{equation}
so once the $k$ integral is performed the energy is proportional to
\begin{equation}
-2\sum_{n=0}^\infty(n+1/2)^3={7\over4}\sum_{n=1}^\infty n^3.
\end{equation}
Thus, the result is 7/4 times the scalar Casimir energy,
\begin{equation}
f_F=-{7\pi^2\over1920 a^4}.
\end{equation}

The Casimir effect so implemented breaks supersymmetry.
However, in a SUSY theory,
if supersymmetric boundary conditions are imposed (the 
unconstrained components of the fermion and scalar fields are all
periodic, for example), the fermionic Casimir energy will just cancel that
due to the bosons.  (For a simple example of how this works, see 
Ref.~\cite{odintsov}.)

\subsection{Casimir Effect on a $D$-dimensional Sphere}

Because of the rather mysterious dependence of the sign and magnitude
of the Casimir stress on the topology and dimensionality of the
bounding geometry, we have carried out a calculation of TE and TM modes bounded
by a spherical shell in $D$ spatial dimensions.  We first consider massless
scalar modes satisfying Dirichlet boundary conditions on the surface, which are
equivalent to electromagnetic TE modes.  Again we calculate the 
vacuum expectation value of the stress on the surface from the Green's 
function.

The Green's function $G({\bf x},t;{\bf x}',t')$,
Eq.~(\ref{vevgreen}),  satisfies the inhomogeneous 
Klein-Gordon equation, or
\begin{equation}
\left ({{\partial^2}\over{\partial t^2}}-\nabla^2\right ) G({\bf
x},t;{\bf x}',t') =\delta^{(D)} ({\bf x}-{\bf x}')\delta (t-t'),
\label{4}
\end{equation}
where $\nabla^2$ is the Laplacian in $D$ dimensions.
We solve the above Green's function equation by dividing space
into two regions,  the interior of a sphere of radius $a$ and
 the exterior of the sphere. On the sphere we impose
Dirichlet boundary conditions
\begin{equation}
G({\bf x},t;{\bf x}',t')\bigm | _{|{\bf x}|=a}=0.
\label{5}
\end{equation}
In addition, in the interior we require that $G$ be finite
at the origin ${\bf x}=0$ and in the exterior we require that $G$
satisfy outgoing-wave boundary conditions at $|{\bf x}|=\infty$,
that is, for a given frequency, $G\sim e^{ikr}/r$.

The radial Casimir force per unit area $f$ on the sphere is
obtained from the radial-radial component of the vacuum expectation 
value of the stress-energy tensor:
\begin{equation}
f =\langle 0|T^{rr}_{\rm in}-T^{rr}_{\rm out}|0\rangle\bigm |_{r=a}.
\label{7}
\end{equation}
To calculate $f$ we exploit the connection between the vacuum expectation
value of the stress-energy tensor $T^{\mu\nu} ({\bf x},t)$ and the Green's
function at equal times $G({\bf x},t;{\bf x}',t)$:
\begin{eqnarray}
f ={1\over 2i}\left[{\partial\over\partial r}{\partial\over\partial r'}G({\bf
x},t;{\bf x}',t)_{\rm in}
-{\partial\over\partial r}{\partial\over\partial r'}
G({\bf x},t;{\bf x}',t)_{\rm out}\right]_{{\bf x}={\bf x}',~|{\bf x}|=a}.
\label{8}
\end{eqnarray}
Adding the interior and the exterior contributions, and performing the
usual imaginary frequency rotation, we obtain the expression for
the stress \cite{bendermilton}:
\begin{eqnarray}
f&=&-\sum_{n=0}^{\infty}w_n(D)
\int_0^{\infty}dx\, x
{d\over dx}\ln\left ( I_{n-1+{D\over 2}}(x)K_{n-1+{D\over2}}(x)x^{2-D}\right ),
\label{e4}\\
&&\qquad w_n(D)={(n-1+{D\over 2})\Gamma (n+D-2)\over
2^{D-1}\pi^{D+1\over 2}a^{D+1} n!\,\Gamma\left ({D-1\over 2}\right
)}.
\end{eqnarray}
It is easy to check that this reduces to the known case at $D=1$, for there
the series truncates---only $n=0$ and 1 contribute, and we easily
find
\begin{equation}
f=-{\pi\over96a^2},
\end{equation}
which agrees with Eq.~(\ref{scalarforce}) for $d=0$ and $a\to2a$.

\begin{figure}
\begin{center}
\epsfig{figure=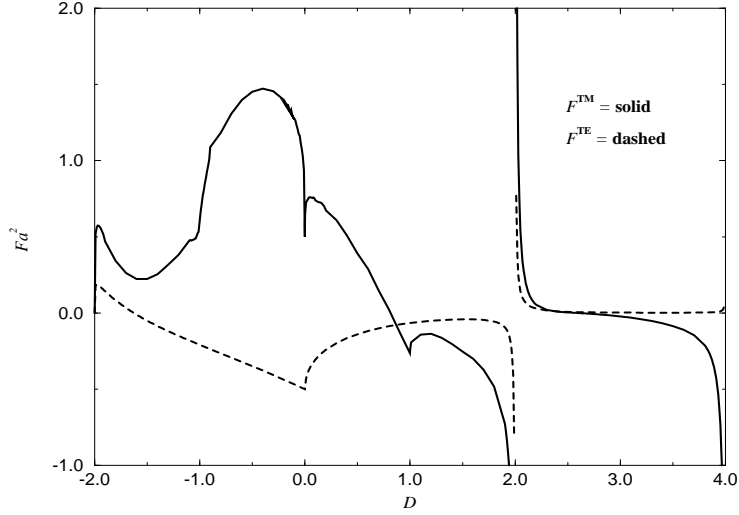,height=4.5in,width=3in,angle=270}
\end{center}
\caption{A plot of the TM and TE (Dirichlet)
 Casimir stress for $-2<D<4$ on a spherical
shell. For $D<2$ ($D<0$) the stress $F^{\rm TM}$ ($F^{\rm TE}$) is complex
and we have plotted the real part.}  
\label{figvdc}
\end{figure}

In general, we proceed as follows:
\begin{itemize}
\item Analytically continue to $D<0$, where the sum (\ref{e4})
converges, although the integrals become complex.
\item Add and subtract the leading asymptotic behavior of the integrals.
\item Continue back to $D>0$, where everything is now finite.
\end{itemize}
The results of numerical evaluations
for the total stress $F$ on the sphere are as shown in Fig.~\ref{figvdc}.


 Note the following features for the scalar modes:
\begin{itemize}
\item Poles occur at $D=2n$, $n=1,2,3,\dots$.
\item Branch points occur
at $D=-2n$, $n=0,1,2,3,\dots$, and the stress is complex for $D<0$.
\item The stress vanishes at negative even integers, $F(-2n)=0$, 
$n=1,2,3,\dots$, but is nonzero at $D=0$: $F(0)=-1/2a^2$.
\item The case of greatest physical interest, $D=3$, has a finite stress,
but one which is much smaller than the corresponding electrodynamic
one: $F(3)=+0.0028168/a^2$. (This result was confirmed in 
Ref.~\cite{lr}.)
\end{itemize}

The TM modes are modes which satisfy mixed boundary conditions on the surface,
\begin{equation}
{\partial\over\partial r}r^{D-2}G({\bf x},t;{\bf x}',t')
\bigg|_{|{\bf x}|=r=a}=0,
\label{tmbc}
\end{equation}
The results are qualitatively similar, and are also shown in Fig.~\ref{figvdc}.
In particular, removing the $n=0$
contribution from the sum of the TE and TM contributions, we recover the
repulsive Boyer result in three dimensions \cite{boyer},
\begin{equation}
E_{\rm sphere}={0.092353\over 2a}.
\end{equation}
For the 3-sphere, the fermionic Casimir energy subject to the 
boundary condition (\ref{bmbc}) is a factor of two
smaller \cite{fermionsp}
\begin{equation}
E_F={0.0204\over a}.
\end{equation}

\subsection{Cylinders}

A similar calculation of the electromagnetic Casimir effect of a perfectly
conducting infinite right circular cylinder 
in three dimensions was performed by De\-Raad and me
twenty years ago.  The calculation is rather more involved, and the 
regularization of the divergences more subtle.  The result for the
Casimir energy per unit length, or the force per unit area, is \cite{dm}
\begin{equation}
{\cal E}=\pi a^2f=-0.01356/a^2.
\end{equation}
Unlike the three-dimensional sphere, the cylinder experiences an attractive
Casimir stress.  Two recent calculations have confirmed this result using
zeta-function techniques \cite{zetacyl1,zetacyl2}.

\section{Force between dielectric slabs}

Over 40 years ago, Lifshitz and collaborators \cite{lifshitz}
worked out the corresponding
forces  between dielectric slabs.  Imagine we have a permittivity which
depends on $z$ as follows:
\begin{equation}
\epsilon(z)=\left\{\begin{array}{cc}
\epsilon_1,&z<0,\\
\epsilon_3,&0<z<a,\\
\epsilon_2,&a<z.
\end{array}\right..
\end{equation}
Then the Lifshitz force between the bodies at zero temperature is given by
($\kappa^2=k^2+\epsilon\zeta^2$, $\zeta$ is the imaginary frequency)
\begin{eqnarray}
f^{T=0}_{\rm Casimir}=-{1\over8\pi^2}\int_0^\infty d\zeta \int_0^\infty
dk^2\,2\kappa_3 \left(d^{-1}+d^{\prime-1}
\right).
\label{dielectricforce}
\end{eqnarray}
Here  the denominators are given by, for the electric (TM) Green's function,
\begin{equation}
d={\kappa_3+\kappa_1\over\kappa_3-\kappa_1}
{\kappa_3+\kappa_2\over\kappa_3-\kappa_2}e^{2\kappa_3 a}-1.
\end{equation}
The magnetic (TE) 
Green's function has the same form as the electric one but with the replacement
\begin{equation}
\kappa\to\kappa/\epsilon\equiv\kappa',
\end{equation}
except in the exponentials; the corresponding denominator is denoted by $d'$.
From this, we can obtain the finite temperature expression immediately
by the substitution
\begin{equation}
\zeta^2\to\zeta_n^2=4\pi^2n^2/\beta^2,
\end{equation}
\begin{equation}
\int_0^\infty {d\zeta\over2\pi}\to{1\over\beta}{\sum_{n=0}^\infty}{}',
\end{equation}
the prime being a reminder to count the $n=0$ term with half weight.
(For a fuller discussion of temperature dependence, see Ref.~\cite{milton}.)

\subsection{ Relation to van der Waals force}
If the central medium is tenuous, $\epsilon-1\ll1$, and is surrounded
by vacuum, for large distances $a\gg \lambda_c$, where $\lambda_c$ is a 
characteristic wavelength of the medium, we can expand the above
general formula and obtain a dispersion-free result:
\begin{equation}
f\approx-{23(\epsilon-1)^2\over640\pi^2a^4}.
\label{longdist}
\end{equation}
For this regime, this should be derivable from the sum of van der Waals
forces, obtained from an intermolecular potential of the form
\begin{equation}
V=-{B\over r^\gamma}.
\end{equation}
We do this by computing the energy ($N= $ density of molecules)
\begin{eqnarray}
E=-{1\over2}B N^2\int_0^a dz\int_0^a dz'\int{(d{\bf r_\perp})(d{\bf
r'_\perp})
\over[({\bf r_\perp-r'_\perp})^2+(z-z')^2]^{\gamma/2}}.
\end{eqnarray}
If we disregard the infinite self-interaction terms (analogous to
dropping the volume energy terms in the Casimir calculation), we
get
\begin{equation}
f=-{\partial\over\partial a}{E\over A}=-{2\pi B
N^2\over(2-\gamma)(3-\gamma)}
{1\over a^{\gamma-3}}.
\end{equation}
So then, upon comparison with (\ref{longdist}), we set $\gamma=7$
and in terms of the polarizability,
\begin{equation}
\alpha={\epsilon-1\over4\pi N},
\end{equation}
we find
\begin{equation}
B={23\over4\pi}\alpha^2,
\label{bee}
\end{equation}
or, equivalently, we recover the Casimir-Polder retarded dispersion 
potential~\cite{casimirpolder},
\begin{equation}
V=-{23\over4\pi}{\alpha^2\over r^7},
\label{caspol}
\end{equation}
whereas for short distances ($a\ll \lambda_c$) we recover
the London potential \cite{london},
\begin{equation}
V=-{3\over\pi}{1\over r^6}\int_0^\infty d\zeta\,\alpha(\zeta)^2.
\end{equation}

Given the divergences of the above calculation, and the
essentially one-dimensional restriction, it is of interest to consider
a tenuous dielectric sphere.  The theory of the Casimir energy for a dielectric
ball was first worked out by me 20 years ago \cite{ball}.
The general expression is of course quite complicated
\begin{equation}
E=-{1\over4\pi a}\int_{-\infty}^\infty dy\,e^{iy\delta}\sum_{l=1}^\infty
(2l+1)x{d\over dx}\ln S_l,
\end{equation}
where
\begin{eqnarray}
S_l=[s_l(x')e'_l(x)-s_l'(x')e_l(x)]^2-\xi^2[s_l(x')e'_l(x)+s_l'(x')e_l(x)]^2,
\end{eqnarray}
where the $s_l$, $e_l$ are spherical Bessel functions of
imaginary argument, the quantity $\xi$ is
\begin{equation}
\xi={\sqrt{\epsilon'\mu\over\epsilon\mu'}-1\over\sqrt{\epsilon'\mu\over
\epsilon\mu'}+1},
\end{equation}
where $\epsilon'$, $\mu'$ represent the permittivity and permeability
in the interior, the corresponding unprimed quantities refer to the exterior,
the time-splitting regularization parameter is  denoted by $\delta$,
and \begin{equation}
x=|y|\sqrt{\epsilon\mu},\quad x'=|y|\sqrt{\epsilon'\mu'}.
\end{equation}
It is easy to check that this result reduces to that for a perfectly
conducting spherical shell \cite{boyer}
 if we set the speed of light inside and out
the same, $\sqrt{\epsilon\mu}=\sqrt{\epsilon'\mu'}$, as well as set
$\xi=1$.  However, if the speed of light is different in the two
regions, the result is no longer finite, but quartically divergent,
and indeed the Schwinger result \cite{js}
follows for that leading divergent term.

Although, in general, this expression
is not finite, there are several methods of isolating the divergences,
at least if the ball is tenuous, ($\epsilon-1\ll1$), and the finite repulsive
observable Casimir energy is \cite{dielectricball}
\begin{equation}
E_{\rm Cas}={23\over1536\pi a}(\epsilon-1)^2.
\label{nobulk}
\end{equation}

What is most remarkable about this result is that it coincides with
the van der Waals energy calculated two years earlier for this nontrivial
geometry.  That is, starting from the Casimir-Polder
potential   (\ref{caspol}) we summed the
pairwise potentials between molecules making up the media.  A sensible
way to regulate this calculation is dimensional continuation, similar to
that described above.  That  is, we evaluate the integral
\begin{equation}
E_{\rm vdW}=-{23\over8\pi}\alpha^2N^2\int d^Dr\,d^Dr'(r^2+r^{\prime2}
-2rr'\cos\theta)^{-\gamma/2},
\end{equation}
where $\theta$ is the angle between $\bf r$ and $\bf r'$, by first
regarding $D>\gamma$ so the integral exists.  The integral may be
done exactly in terms of gamma functions, which when continued to
$D=3$, $\gamma=7$ yields
Eq.~(\ref{nobulk}) \cite{mng}.

Thus there can hardly be any doubt that the Casimir effect,
in the tenuous limit, coincides with the van der Waals attraction between
mole\-cules.  This seems to go some way toward providing understanding of
this zero-point fluctuation phenomenon.  But the subject is not closed.
Two years ago Romeo and I demonstrated that for a dilute cylinder
the van der Waals energy is {\em zero} \cite{vdwc,zetacyl2}.  
Presumably the same holds for the Casimir energy, in order $(\epsilon-1)^2$,
but a demonstration of that is not yet at hand. (It turns out to be quite
difficult to compute the Casimir effect for a dielectric cylinder.)
 Remarkably,
for a dilute cylinder with constant speed of light inside and out,
$\epsilon\mu=\mbox{constant}$, it has been demonstrated that the Casimir
energy vanishes, that is, the energy is of order
$\xi^4$ \cite{zetacyl2,zetacyl3},
 but that would seem to be a completely different case.

\section{``Dynamical Casimir effect'' and relevance to sonoluminescence}

{\em Sonoluminescence\/} refers to that remarkable
phenomenon in which a small bubble of air injected into a container of
water and suspended in a node of a strong acoustic standing wave emits
light.  More precisely, if it is driven with a standing wave of about
20,000 Hz at an overpressure of about 1 atm, the bubble expands and contracts
in concert with the wave, from a maximum radius $\sim 10^{-3}$ cm to a 
minimum radius of $\sim 10^{-4}$ cm.  Exactly at minimum radius roughly
1 million optical photons are emitted, for a total energy liberated of 10 MeV.
For a review of the experimental situation as of a few years ago, see
Ref.~\cite{review}.

Julian Schwinger, informed of these experiments by Putterman, immediately
assumed the effect derived from a dynamical version of the Casimir effect,
and published in the last few years of his life a series of papers in the
PNAS-USA attempting to account for the observations in this way \cite{jss}.
 Subsequently
a number of other have jumped on this bandwagon \cite{band,carlson,sciama}.

The problem is that the ``dynamical Casimir effect'' remains largely 
unknown.\footnote{Of course, radiation from a moving mirror is well
studied \cite{mirror}. For cavities, little beyond perturbation theory,
valid for ``small but arbitrary dynamical changes,''
hardly relevant to the profound changes seen in sonoluminescence, is
known.  For recent references  see Ref.~\cite{soff}.}
Schwinger and his followers had to rely on the known results for the
Casimir effect with static boundary conditions.  Two possible avenues then
appeared:
\begin{itemize}
\item One could employ the adiabatic approximation, which does not seem
unreasonable, since the time scale for emission of the photons, 
$\sim 10^{-11}$~s, 
is far longer than the characteristic optical time scale, $\sim 10^{-15}$ s.
Schwinger \cite{jss}, and Carlson et al.~\cite{carlson}, 
then used the divergent bulk energy term
\begin{equation}
E_{\rm bulk}={4\pi a^3\over3}
\int{(d{\bf k})\over(2\pi)^3}{1\over2}k\left(1-
{1\over n}\right).\label{bulk}
\end{equation}  If a reasonable cutoff is inserted, the required
10 MeV of energy is indeed present.
\end{itemize}

My objection to this is that it now seems unequivocal that the Casimir
energy for a (dilute) dielectric ball is given by the finite expression
(\ref{nobulk}). The divergent expression (\ref{bulk}) is properly absorbed in
a renormalization of the material properties.  If Eq.~(\ref{nobulk}) is used,
the Casimir energy is {\em
 ten orders of magnitude too small to be relevant.}

\begin{itemize}
\item The second avenue, followed by Schwinger \cite{jss} and Liberati et 
al.~\cite{sciama}, is to use
the instantaneous or sudden approximation, in which  the static bubble simply
disappears.  Then the photon production rate can be calculated from the
overlap of the two static configurations, through the Bogoliubov coefficients.
Again, reasonable agreement with  observations is reported.
\end{itemize}

My objection here is that impossibly short time scales are required.
For the sudden approximation to make sense, the time scale $\tau$ 
of collapse must
be small compared to $10^{-15}$ s.  But in fact, the overall collapse time
scale is $10^{-4}$ s, and the emission time scale is probably $10^{-11}$ s.
An estimate of what one expects with reasonable numbers may be obtained
from the Unruh temperature \cite{unruh}
\begin{equation}
T={A\over2\pi},
\end{equation}
where $A$  is  the acceleration of the surface.
Estimating $A$ by $a/\tau^2$, $a$ being some relevant bubble radius,
 and recognizing that the observations are
consistent with a photon temperature of order 20,000 K, we estimate the
required time scale by
\begin{eqnarray}
\tau^2\sim{a\over2\pi T}{\hbar\over c}\sim {10^{-4}
 \mbox{cm} \times 2\times10^{-5}
\mbox{eV\,cm}\over(10 \mbox{ eV}) (3\times10^{10} \mbox{cm/s})^2} 
\sim10^{-31}\,\mbox{s},\quad \tau\sim 10^{-15}\,\mbox{s},
\end{eqnarray}
which once again seems physically unrealizable.\footnote{The same conclusion
is reached if the Larmor formula for the radiated power, $P=(2/3)(e^2/c^3)
A^2$, the leading approximation in Ref.~\cite{soff}, is used.  See also
Ref.~\cite{mng1}}

\section{Conclusions}
A great many isolated facts about the Casimir effect are now known:
\begin{itemize}
\item The dimensional dependence for planes\footnote{Although results
have been given for rectangular cavities, for example by \cite{aw},
these results include only interior modes, and are thus suspect.}
 and hyperspheres is now known.
\item The equivalence of the van der Waals force and the Casimir force
for dilute media is definitively established.
\item Moreover, important applications of the Casimir effect to many fields,
such as cosmology,\footnote{A recent example is Ref.~\cite{odintsov}.} 
have been made.
\end{itemize}

However, we have very little understanding of the effects.
\begin{itemize}
\item We cannot predict, {\it a priori}, the sign of the effect.
\item There are divergences occurring, for example with the dielectric ball,
the nature of which is not well understood.
\item Physically, why does the Casimir energy for a sphere in even dimensions
diverge?
\item Most importantly, can one make progress in understanding the dynamical
Casimir effect?
\end{itemize}

\noindent{\bf Acknowledgements.} This work was supported in part by
the US Department of Energy.  I am grateful to the organizers of the
Fradkin Memorial Conference for allowing me to make this presentation in
honor of the many great works of Efim Fradkin.  I thank Peter van
Nieuwenhuizen and Gerhard Soff for helpful discussions at the conference.


\end{document}